\documentclass[a4,12pt]{article}
\usepackage[utf8]{inputenc}
\usepackage{mathenv}
\usepackage{color}
\usepackage{amsfonts}
\usepackage{calrsfs}
\usepackage{amsmath}
\usepackage[T1]{fontenc}
\usepackage[french,english]{babel} 
\usepackage{graphicx}
\usepackage{enumitem}
\usepackage{mathabx}
\usepackage{placeins} 
\usepackage{subfig}
\usepackage{version}
\usepackage{bbold} 
\usepackage{pifont}
\usepackage{qtree}
\usepackage{stmaryrd} 

\usepackage[pdftex,hmargin={2cm,2cm},vmargin={2cm,2cm}]{geometry}  

\usepackage[pdftex,bookmarks=false,citecolor=blue,colorlinks=true,urlcolor=red]{hyperref} 

\usepackage{bm}
\def\N{\mathbb N}

\usepackage{setspace}

\usepackage{caption}
\graphicspath{{Figure/}}

\title{Episome partitioning and symmetric cell divisions: quantifying the role of random events in the persistence of HPV infections}

\author{Thomas Bénéteau\up{$\star$}, Chrisian Selinger\up{+}, Mircea T. Sofonea\up{+}, \\Samuel Alizon\up{+}}

\date{\small MIVEGEC, Univ Montpellier, CNRS, IRD, Montpellier, France \\
$^\star$ Corresponding author: thomas.beneteau@ird.fr \\
$^+$ equal contribution
}

\begin{document}


\maketitle

\begin{abstract}
Human Papillomaviruses (HPV) are one of the most prevalent sexually transmitted infections (STI) and the most oncogenic viruses known to humans. The vast majority of HPV infections clear in less than 3 years, but the underlying mechanisms, especially the involvement of the immune response, are still poorly known. Building on earlier work stressing the potential importance of stochasticity in HPV infection clearance, we develop a stochastic mathematical model of HPV dynamics with an explicit description of the intracellular level.
We show that the random partitioning of virus episomes upon stem cell division and the occurrence of symmetric divisions greatly affect viral persistence. These results call for more detailed within-host studies to better understand the relative importance of stochasticity and immunity in HPV infection clearance.
\end{abstract}

\subsection*{Authors summary}

Every year, infections by Human Papillomaviruses (HPV) are responsible for a large share of infectious cancers. Most HPV infections (80 to 90\%) are cleared naturally within three years. Among the few that persist into chronic infections, the majority (88\%) also regress. Hence for a given HPV infection, the risk of progression towards cancerous status is low. Unfortunately, the prevalence of HPVs is very high, which makes it a major public health issue. The immune response is often invoked to explain HPV clearance in non-persisting infections but many uncertainties remain. Besides immunity, randomness has also suggested to play an important role.
Here, we examine how random events occurring during the life cycle of the virus could alter the persistence of the virus inside the host. We develop a mechanistic model that explicitly follows the dynamic of viral copies inside host cells, as well as the dynamics of the epithelium. In our model, infection extinction can occur because all the virus copies end up in the differentiated cell upon cellular division or because a stem cell divides symmetrically to generate two differentiated cells. 
We find that the combination of these random events drastically affects infection persistence. More generally, the importance of these random fluctuations could match that of immunity and call for further studies at the within-host and at the epidemiological level.  

\section{Introduction}

Human Papillomaviruses (HPV) are the most oncogenic viruses known to humans \cite{Plummer2016}. Up to 8.6\% of all cancers occurring worldwide in women are caused by HPV infections, versus  0.8\%~for cancers in men \cite{Martel2017}. Among the various types of cancers attributable to HPV, cervical cancers, for which the causative effect of the virus has long been established \cite{Bosch2002,Leemans2011, Crow2012, Martel2017}, stand out since they are the third-most diagnosed form of cancer in women \cite{Schiffman2013}.

The burden HPVs impose on human populations largely originates from their high prevalence in the general population. For example, it was estimated that the average lifetime probability (from sexual debut to age 70) of acquiring a genital HPV infection in the US was 84.6\% for women and 91.3\% for men \cite{Chesson2014}. But since up to 90\% of these infections remain subclinical and are cleared in less than 3 years \cite{Schiffman2011}, viral persistence, in particular for certain genotypes such as HPV-16, determines potential oncogenicity \cite{Insinga2007}.

Despite its ubiquity, the mechanisms underlying HPV infection clearance and persistence are still poorly understood. Although the innate immune response (e.g. natural killer cells) is known to be involved in chronic infections \cite{VanHede2014}, few data exist for non-persistent infections. Adaptive immune responses (e.g. the role of T helper cells) have been studied in the context of vaccine trials, but the results remain inconsistent~\cite{Scott2001}. It has been noticed that people with recurrent genital warts were more likely to be HPV seropositive than those who do not have a history of genital warts \cite{Carter1995}. However, more generally, the breadth and magnitude of adaptive immune responses following HPV infection are believed to be limited \cite{Stanley2008} due to innate immune evasion mechanisms and a lack of inflammatory responses elicited by infected epithelial cells. Furthermore, in natural infections, there is no viremia (\textit{i.e.} presence of viruses in the blood) and the production of detectable antibodies against the virus is low \cite{Stanley2008}. Another longitudinal study investigated HPV-16 seroconversion statuses among various groups of infected women but did not reach any conclusive result, potentially by lack of statistical power \cite{Farhat2009}. 

Finally, the inherently stochastic nature of the episomal life cycle of HPV could be an additional factor contributing to the clearance of HPV infections. HPVs are non-lytic viruses that maintain in the basal stem cells by replicating with host cell chromosomes \cite{Moody2017}. Infection persistence in the epithelial tissue occurs when viral copies are transmitted to the daughter stem cell during cell divisions, viral copies in the daughter differentiated cell migrate and maturate with the cell, before being released in the surface upon shedding.
In particular, the potential symmetric division of infected basal stem cells into two daughter stem cells or two differentiated cells \cite{Clayton2007, Doupe2012} clearly impacts the spread and persistence of the virus in the basal layer of the tissue. As HPV infections begin with few infected cells \cite{Doorbar2006}, randomness in the type of cell division plays a crucial role in the clearance of HPV infections \cite{Ryser2015}. 
A second source of stochasticity occurs within the nucleus of infected cells, where the HPV genome replicates as independent circular DNA copies called episomes. During stem cell divisions the number of episomes that are passed on to each daughter cell is random. 
Since cells are infected with a small number (potentially unique) of viral copies \cite{Doorbar2012}, stochasticity in episome distribution upon division could impact the persistence of HPV infection. The partition of viral plasmids in human cells is uncommon but not specific to HPV. This phenomenon is still not clearly understood despite its importance in the maintenance of infections and progression towards cancers \cite{Chiu2018}. To date, this aspect of HPV infections has not been taken into account. We integrate this aspect in our work combined with the randomness in cell divisions symmetry.

Taken separately, the intra- and inter-cellular stochastic processes fail to capture part of the life cycle of HPV infections. The intracellular approach does not account for the spread of the infection inside the epithelial tissue, whereas the inter-cellular approach ignores the necessity of viral persistence in proliferating cells. Thus, we hypothesise that combining the two approaches can provide additional insights into the dynamics of HPV infection.

We investigate the role of stochasticity in the clearance of HPV infections and quantify the role of both symmetric divisions and episomal partitioning in this process. We provide further evidence for the crucial effect of randomness in the clearance of HPV infections. 

\section{Methods}

Our model focuses on the basal layer of a stratified squamous epithelium (SSE), but it can easily be adapted to other tissues as the HPV replication pattern is essentially the same in most sites \cite{Rautava2012, Stanley2012}. We only follow the dynamics of episomes in infected stem cells because SSE are in perpetual renewal and the turnover intervaine of the non-dividing cells lasts 3 to 6 weeks \cite{Averette1970, Stanley2006, Stanley2010}. Therefore, in the absence of cell-to-cell transmission, it is improbable that an infection could persist only in the differentiated layers of SSE. We also neglect potential re-infections of the SSE from virions released by desquamating cells because these are thought to be unlikely given the importance of micro-wounding in the establishment of new infections \cite{Moody2017}.

The life-cycle is divided into two steps following the non-lytic properties of HPV: 1) during the division of an infected cell, HPV episomes are distributed randomly between the two daughter cells, and 2) infected stem cells can divide symmetrically, thereby ending or, conversely, further spreading the infection. Both the intra- and inter-cellular mechanisms are modelled as discrete-time branching processes \cite{Harris1948,Athreya1972}.

More formally, let $X_n, n\in\mathbb{N}$ be the total number of HPV episomes in all infected stem cells, $n$ being the number of stem cell generations since the beginning of the infection. Unless stated otherwise, a time step corresponds to one cell generation.  
$X_n$ is a random variable, the distribution of which is non-trivial, for it is constructed following several random events. We assume the random events between cell lineages, \textit{i.e.}~descending cells by asymmetric divisions, to be independent and identically distributed. Within a cell lineage, we assume other stochastic events to be identically distributed (but not independent). Independence is verified under certain specific conditions, for instance if we assume a cell lineage only divides asymmetrically.

\subsection{Intracellular dynamics: viral distribution during cell division}

\label{section:intra_cell_dyn}

HPVs are non-lytic viruses that follow the life cycle of their host cells. In basal (stem) cells, the number of viral copies is usually considered to be limited to several dozens of episomes per cell: 10 to 200 according to \cite{Doorbar2006}, and 50 to 100 according to \cite{Moody2017}. We denote by $C$ the maximum number of episomes per infected stem cell. We model intracellular dynamics as the superposition of two distinct random steps: 
\begin{enumerate}[label=(\roman*)]
    \item A division, where each episome in the dividing cell is distributed to the two daughter cells according to a Bernoulli distribution $Bernoulli(p)$. For asymmetric divisions, parameter $p$ stands for the probability for an episome to be relocated in the daughter stem cell. Episomes distributed to the differentiated daughter cell are considered withdrawn from the system as they migrate to the surface without contributing to the infection of stem cells.
    
    \item An amplification, which occurs once the division over. For this random phenomenon we either assume a Dirac distribution $\delta(\lambda)$ or a Poisson distribution $Poisson(\lambda))$, with $\lambda\in\N$. We refer to $\lambda$ as the episome replication factor.
\end{enumerate}

The order of the two steps can be reversed without affecting the intrinsic characteristics of this branching process (see appendix \ref{appendix:amplification_before_partition} for more details). Since HPV genomes are present in low copy numbers and HPV gene expression is set at a minimum level in basal stem cells \cite{Stanley2012}, we assume that the number of HPV viral copies does not affect the dynamics of the host stem cells.

The infection can spread in the basal layer following symmetric divisions in two stem cells, a phenomenon governed by strong stochastic effects and detailed in the next section. To distinguish between daughter stem cells, we use the notation described in appendix \ref{appendix:annotation} and define $p$ as the probability for an episome to be distributed in the left (\textbf{L}) daughter stem cell. We assume episome distribution bias towards one of the daughter cells does not apply when the two daughter cells are stem cells, hence we assume distribution to be even ($p=0.5$) during symmetric division.

\subsection{Intra-host episome dynamics}

In what follows, we group all the layers apical to the basal layer in a general population of differentiated cells. Let $(\Gamma_n)_{n\in\N}$ be the number of infected stem cells at time $n$. We denote by $s$ resp. $r$ the probability of a symmetric division to yield two stem cells resp. two differentiated cells. We assume $s$ and $r$ to be small compared to the probability of asymmetric cell division. Following \cite{Ryser2015}, we model infected stem cells dynamics using the following branching process:

\begin{align}
    \Gamma_{n+1} = \sum \limits_{i = 1}^{\Gamma_n} \phi_i 
\end{align}
where $\phi_i$ are independent and identically distributed random variables with the following distribution:
\begin{align}
    \phi_i = \left\{ \begin{array}{ll}
        2  &  \text{with probability $s$}\\
        0  &  \text{with probability $r + \psi$}\\
        1  &  \text{with probability $1-r-s-\psi$}
    \end{array}
    \right.
    \label{eq:branching_general}
\end{align}

where $\psi$ is a random variable accounting for intracellular effects leading to the end of the infection of the infected stem cell $S$.  A histological study from 1970 estimated that $r+s\leq 0.08$ \cite{Averette1970}, and a more recent analysis also estimated 8\% of cell divisions to be symmetric \cite{Clayton2007}.

\subsection{Numerical simulations and sensitivity analysis}

\begin{table}[ht!]
    \centering
    \begin{tabular}{p{2cm}p{5cm}p{2cm}p{3cm}}
        \textbf{Notation} & \begin{center} \textbf{Description} \end{center}  & \textbf{Range} & \begin{center} \textbf{Source} \end{center} \\ \hline \hline
        $p$ & Probability for an episome to be distributed to the daughter stem cell& [0, 1] & \cite{McPhilips2005} \\ \hline
        $\lambda$ & Episome replication factor (number of viral copies) & $\geq 2$ &  --- \\ \hline
        $s$ & Probability of symmetric divisions into two stem cells & [0.01, 0.04] & \cite{Averette1970, Clayton2007} \\ \hline
        $r$ & Probability of symmetric divisions into two differentiated cells & [0.01, 0.04] & \cite{Averette1970, Clayton2007} \\ \hline
        $C$ & Viral capacity per infected stem cell & [10, 200] & \cite{Doorbar2006, Moody2017} \\ \hline
        $N_0$ & Number of episome in each infected cells at the beginning of the epidemic & [1, C] & --- \\ \hline
        $\Gamma_0$ & Number of infected stem cells at the beginning of the epidemic & $\geq 1$ & ---\\ \hline
    \end{tabular} 
    \caption{\textbf{Numerical parameters used in the model.} The parameters $p, \lambda, C$ and $N_0$ govern the intracellular process, the three others are involved in the inter-cellular approach. Another parameter (not numerical) can be added to the list: the type of intracellular amplification (Dirac or Poisson distributed) as the two types slightly differ  in their construction (for further details, see Section \ref{section:intra_cell_dyn} and Appendices \ref{appendix:intra_process}, \ref{appendix:amplification_before_partition}).}
    \label{tab:Notation}
\end{table}
\FloatBarrier

Some analytical insights can be obtained from the previous equations but in-depth investigations require numerical simulations. We estimated the cumulative probability of extinction ($p_{ext}$) given a set of parameters (see Table \ref{tab:Notation} for further details on the parameters used) by carrying out $\Omega=10^3$ independent runs parametrised by the same parameter values. We also include the type of intracellular amplification in the analysis (Dirac or Poisson distributed, see Section \ref{section:intra_cell_dyn}). Using the $\Omega$ different trajectories, we calculate the cumulative probability of extinction $p_{ext}(t)$ as follows:

\begin{align}
    p_{ext}(t, \mathcal{S}) = \frac{1}{\Omega} \sum \limits_{k=1}^\Omega \mathbb{1}_{\{X^{k}_t = 0|\mathcal{S}\}}
\end{align}

where $X^{k}_t|\mathcal{S}$ is the total number of episomes at time $t$ given parameters set $\mathcal{S}$ for the $k$-th trajectories. If $\Gamma_0 >1$, we assume that all infected cells initially harbour $N_0$ viral copies, hence a total of $X_0 = \Gamma_0\times N_0$ episomes.
$p_{ext}$ is computed for all $t\leq 100$ cell generations. Based on data from the 1970s \cite{Averette1970}, we bound the rate of divisions between $[0.03, 0.07]$ day$^{-1}$, which means the upper bound of 100 cell divisions is sufficient to estimate infection clearance after $3$ years. For each $t\in\N$, we estimate $p_\text{ext}$ using Maximum Likelihood estimation for a Binomial proportion. For each of these proportion we computed the 95\% confidence interval using Agresti-Coull method \cite{Agresti1998}.

\bigskip

To evaluate the role of each parameter on the probability of extinction, we perform a global sensitivity analysis using variance-based methods (also known as Sobol indices). This approach decomposes the variance of the simulation outputs (here $p_\text{ext}$) into fractions that can be attributed to each of our parameters or interactions between them \cite{Sobol2001, Saltelli2007}. To do so, we store parameters sets in two different matrices $A$ and $B$  of dimension $n\times 7$, where the columns correspond to the six parameters $p,\lambda,s,r,C$, and $N_0$ (Table~\ref{tab:Notation}) and the type of intracellular amplification. We denote by $A_B^{(k)}$ the matrix $A$ whose $k$-th column has been replaced by the $k$-th column of matrix $B$. We generate $A$ and $B$ sets using the Latin Hypercube Sampling (LHS) method implemented in the  \textsf{lhs} package \cite{lhsR} in R v4.0.3 \cite{R4.0.3}. Each matrix contains $n=500$ different parameter sets, see appendix \ref{appendix:LHS} for more details. We execute the global sensitivity analysis using \textsf{multisensi} package \cite{multisensipckg}. We computed model outputs (i.e. extinction probabilities) for all sets $A$, $B$, $A_B^{(k)}, k\in\llbracket1, 7\rrbracket$ \cite{Saltelli2010} to obtain the following Monte-Carlo estimator for the variance of the extinction probability up to time $t$:
\begin{equation*}
    \text{Var}_i(t) = \frac{1}{\Omega}\sum_{k=1}^{\Omega} p_{ext}(t,B)_k \left( p_{ext}(t,A_B^{(k)})-p_{ext}(t,(A)_k) \right)
\end{equation*}

 We do not include the initial number of infected stem cells ($\Gamma_0$) in this analysis, since the sensitivity of $p_{\text{ext}}$ with respect to $\Gamma_0$ follows from the branching process property: 
\begin{align}
    \mathbb{P}[X_t = 0|\Gamma_0 = \nu] = \mathbb{P}[X_t = 0|\Gamma_0 = 1]^\nu
\end{align}

\section{Results}

\begin{figure}[ht]
    \centering
    \includegraphics[scale=0.8]{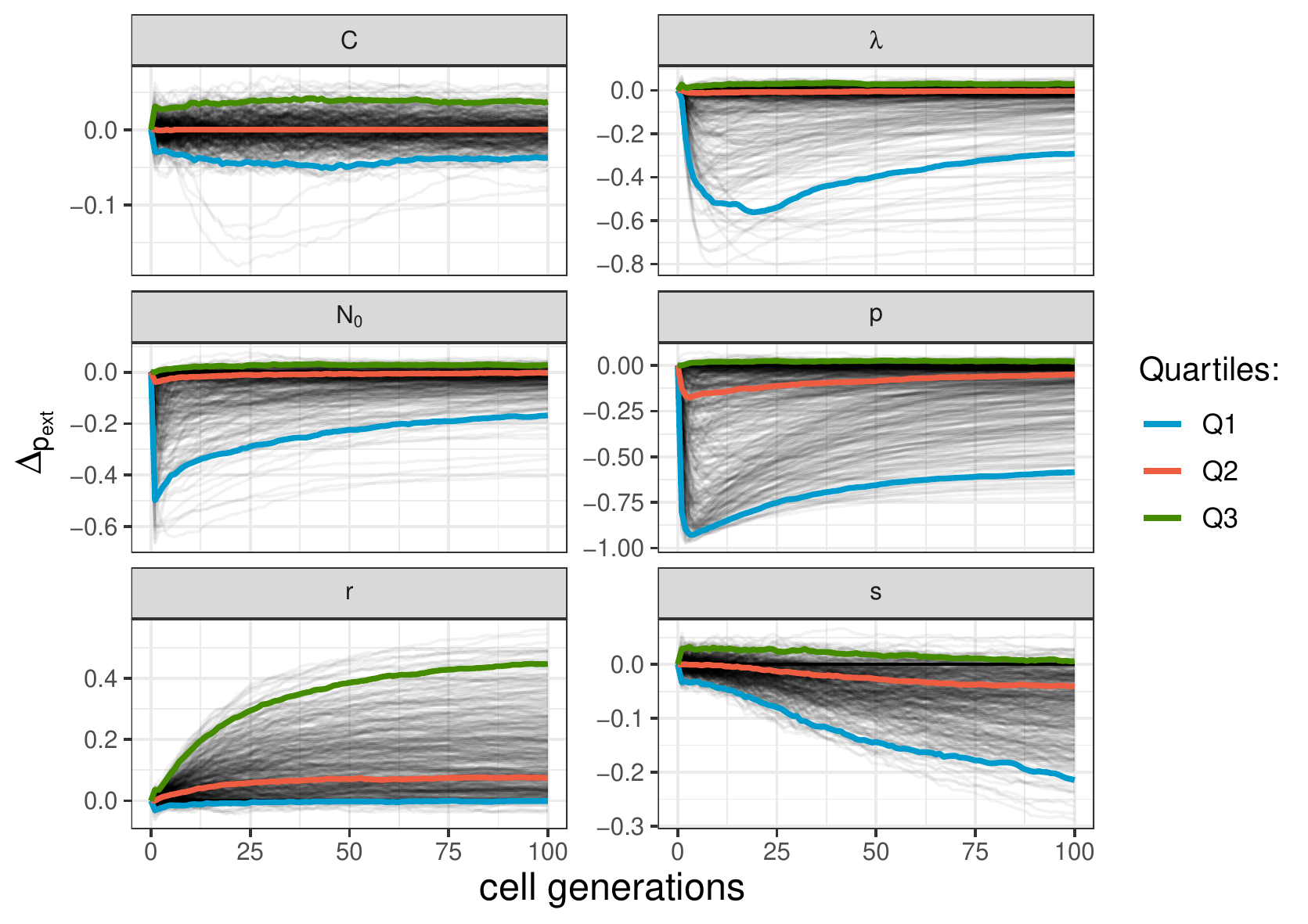}
    \caption{\textbf{Difference in the cumulative probability of extinction for two sets of parameters that only differ by one value.} On each panel, the title indicates which parameter varies between the two sets. In each of this panel, we subtract the probability of extinction for the lowest of the two varying parameters to the probability of extinction of the highest of the two.
    We find consistent results as theory prediction: the probability of extinction decreases with $\lambda, p, s, N_0$ and increase with $r$. The viral capacity in cell ($C$), has little to no effect on the probability of infection. Line colors show the median (in red) and the first and third quartiles (in blue and green). }
    \label{fig:variation_1D}
\end{figure}

Our model contains two nested branching process, which makes it difficult to obtain analytical results. However, analytical expressions can be derived if we neglect either of the two stochastic processes (see section \ref{appendix:intra_process} for the branching process without inter-cellular aspect and section \ref{appendix:inter_process} for the branching process without intracellular stochasticity) and we use these results as boundaries to interpret the results of our numerical simulations and to determine how the cumulative probability of extinction ($p_\text{ext}$) varies with the different parameters. Simulations also allow us to explicit this direction of variation in $p_\text{ext}$ with each parameter. Figure \ref{fig:variation_1D} displays the difference between probabilities of extinction for two parameter sets that only differ by the value of one parameter, we subtract the probability of extinction for the lowest of the two varying parameters to the probability of extinction of the highest of the two.

\FloatBarrier
\subsubsection*{Intracellular stochasticity}

\begin{figure}[t!]
    \centering
    \includegraphics{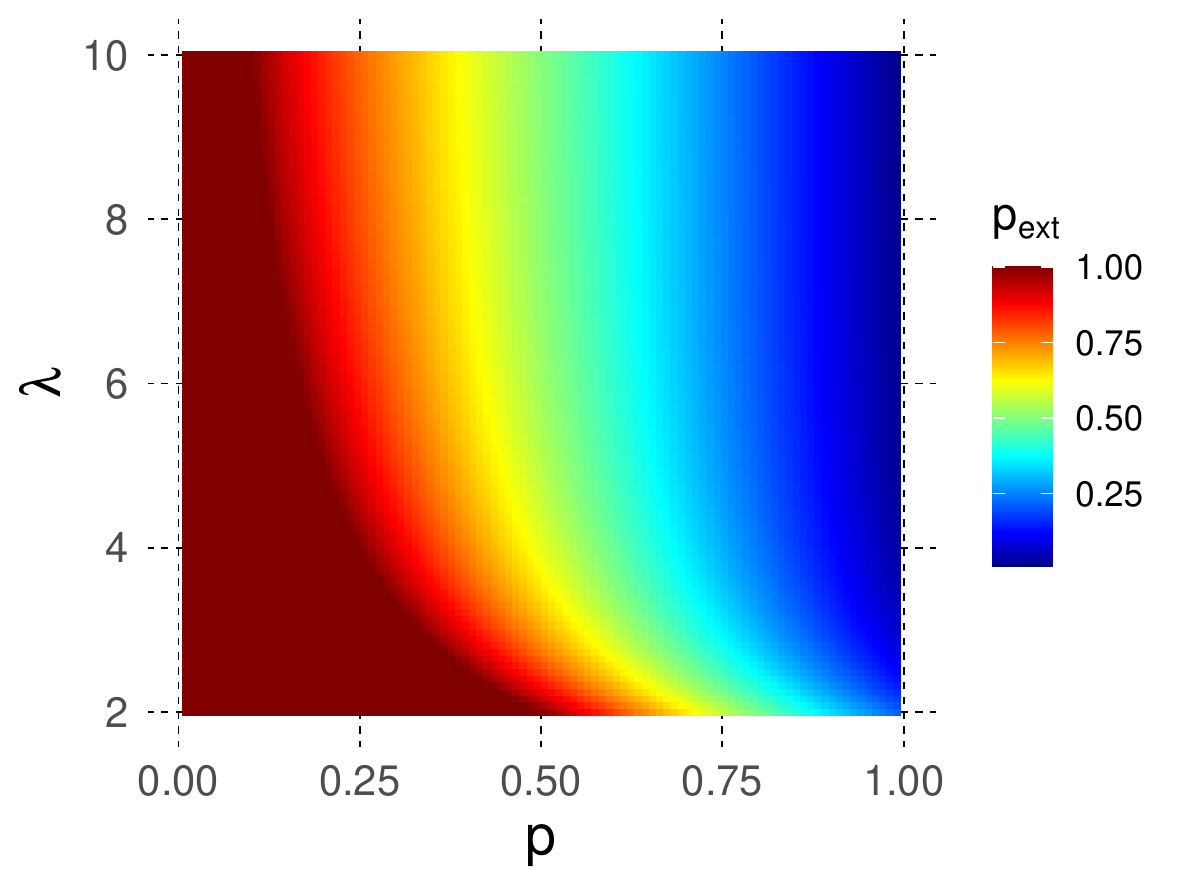}
    \caption{\textbf{Effect of episome replication ($\lambda$) and asymmetry in episome partitioning ($p$) on the cumulative probability of extinction ($p_{\text{ext}}$).} We here assume a Poisson scenario. $p_{\text{ext}}$ decreases with both $\lambda$ and $p$, and reaches a threshold $p_{\text{ext}}=1-p$ for a given $p$ when $\lambda$ increases. This is consistent with the fact that if $\lambda$ is sufficiently high, the main source of extinction is the first cell division of a stem cell containing 1 episome.
    }
    \label{fig:pext_poiss_post}
\end{figure}

As detailed in Appendix \ref{appendix:intra_process}, if stochasticity only acts at the within-cell level, \textit{i.e.}~symmetric divisions are assumed not to occur, we retrieve classical results from Bienaymé-Galton-Watson (BGW) branching process theory \cite{Athreya1972}: the extinction probability $p_{\text{ext}}$ then decreases with the episome replication factor, $\lambda$, and with the probability for an episome to end up in the daughter stem cell, $p$ (Figure \ref{fig:pext_poiss_post}).  As $\lambda$ increases, we observe that $p_{\text{ext}}$ converges toward $1-p$. Biologically, this means that the main event governing the probability of extinction is the first cell division and the distribution of the original episome into a differentiated or a stem cell. 

\begin{figure}[t!]
    \centering
    \includegraphics[scale=0.7]{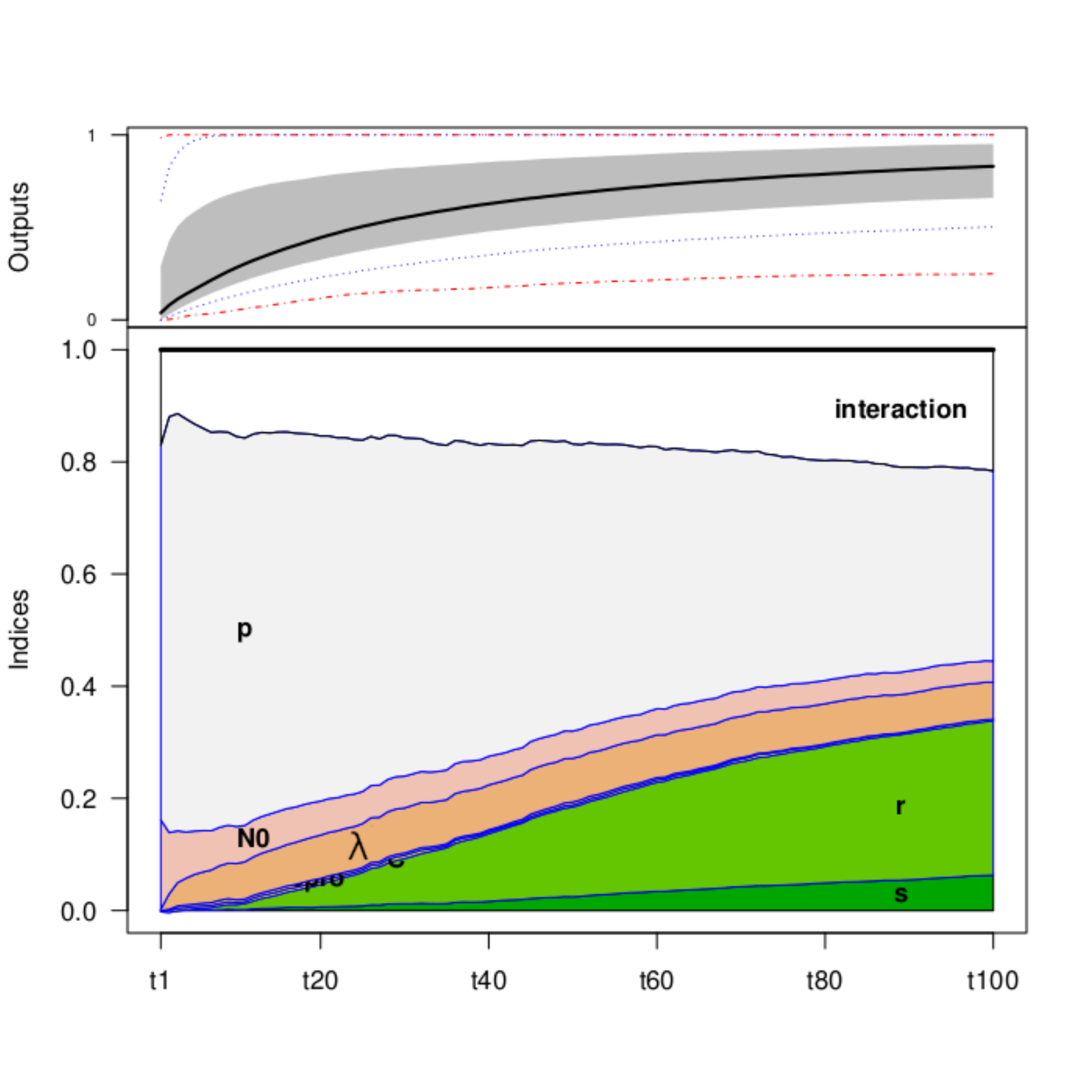}
    \caption{\textbf{Variance-based sensitivity analysis for the cumulative probability of extinction ($p_\text{ext}$).} The upper subplot shows the extreme (dashed line), inter-quartile (grey) and median (dark) probabilities of extinction estimated for the $2$ sets of $500$ parameters. The lower subplot displays the cumulated first-order sensitivity indexes ($\text{Var}_i/\text{Var}_{p_\text{ext}}$, with $\text{Var}_{p_\text{ext}}$ the total variance of $p_\text{ext}$) for all time steps and all parameters. intracellular processes dominate in the early steps of the infection. The importance of inter-cellular processes increases over time, mostly through the probability of symmetric divisions in two differentiated cells. Intracellular viral capacity $C$ and type of reproduction (e.g. fixed or Poisson, see Section \ref{section:intra_cell_dyn}) are barely visible due to their small effect.} 
    \label{fig:Sobol_norm}
\end{figure}

Numerical simulations allow us to investigate the effect of intracellular processes while accounting for symmetric divisions. In particular, our sensitivity analysis highlights the effect of $\lambda$ and $p$ (Figure \ref{fig:Sobol_norm}). At the beginning of the infection, intracellular phenomenon explains around 70\% of the variance, mostly carried by the probability of allocation to daughter stem cell $p$. Indeed, in the early phase of the infection, the total number of episomes is usually low and, since symmetric divisions are rare ($r+s \leq 0.08$\cite{Averette1970}), the early dynamic of episomes is mostly governed by intracellular process. We also retrieve the results from the analytical model, although the decrease in $p_{\text{ext}}$ with increasing $\lambda$ is more limited (Figure \ref{fig:variation_1D}).

As explained in the methods, the numerical simulations slightly deviate from the analytical framework because the number of episomes per cell is assumed to be limited to $C$. However, we expect most of these results to hold in the general framework because we focus on a maximum time scale of 100 cell generations. We observe more deviations on that time interval when $C$ is small and when the intracellular regime is slightly supercritical (see appendix \ref{appendix:intra_process}).

\subsubsection*{Inter-cellular stochasticity}

When omitting intracellular stochasticity, theory allows us to predict that the branching process will go extinct with probability $p_{\text{ext}}=\min(r/s, 1)$, which we confirm with numerical simulations (see Appendix \ref{appendix:inter_process} for more details). Besides, we also find consistent results with theory as the estimated $p\text{ext}$ decreases (resp. increases) with $s$ (resp. $r$), see Figure \ref{fig:variation_1D}.

As intracellular stochastic processes only increase the risk for the infection to go extinct, we can lower bound the cumulative probability of extinction of our general process by the cumulative probability of extinction for the sole inter-cellular a (see Appendix \ref{appendix:inter_process}). Generally speaking, no matter how rapidly the virus replicates at the intracellular level, its spread is bounded by the cellular dynamics of the host tissue. Therefore, we have 
\begin{align}
    \mathbb{P}[X_\infty=0|\Gamma_0=1] \geq \min\left(\frac{r}{s}, 1\right) 
\end{align}

Additionally we can notice in equation \ref{eq:branching_general} that if the inter-cellular process is critical (i.e. $r=s$), then the general process is subcritical. Indeed, under such circumstances we have 
$$m:=\mathbb{E}[\phi] = 1-\psi \leq 1 \Rightarrow p_\text{ext} = 1$$.

In Figure \ref{fig:Sobol_norm}, the proportion of the variance explained by symmetric divisions increases over time. Indeed, on average the first symmetric division occurs after $(1/(r+s)$ cell generations. Hence, the early phase of the infection is mostly governed by intracellular aspect. But later in the infection, if the  time between two symmetric events is sufficiently large, the intracellular dynamic has almost surely converged to one of the two states $\{0, C\}$. Under such circumstances, only the inter-cellular is acting on the dynamic of the infection. We observe a significant difference in the proportion of variance explained by the parameter $r$ compared with $s$, the latter having a smaller slope. This difference is directly induced by the intracellular stochasticity that can terminate infections prematurely.

\FloatBarrier
\subsubsection*{Initial conditions}

We then assess the effect on the probability of extinction of the initial number of episomes and their distribution in one or more stem cells. Let us assume the branching process to start with a total of $X_0 = \nu \times \omega$ episomes, $(\nu, \omega)\in\mathbb{N}^2, \nu>\omega$. We then look for the optimal distribution of these episomes from the virus' perspective. For this, we evaluate the probability of extinction under two extreme scenarios: i) $\nu$ cells each containing $\omega$ episomes, or ii) $\omega$ cells each containing $\nu$ viral copies. From branching theory, we know that for the inter-cellular level, the probability of extinction when starting with $\alpha\in\N$ identical and independent infected cells is equal to the probability of extinction of one cell raised to the power $\alpha$.

Therefore, we only need to compute the probabilities of extinction for two sets of parameters that are identical, except for the number of episomes and number of infected stem cells at time $t=0$, that are switched. We use numerical simulations to evaluate those probabilities. Figure \ref{fig:pair_nu} shows the estimations of $p_\text{ext}$ in the two extreme for six nearly-identical sets of parameters that only vary with respect to $N_0$ and $\Gamma_0$.

\begin{figure}[ht]
    \centering
    \includegraphics[scale = 0.8]{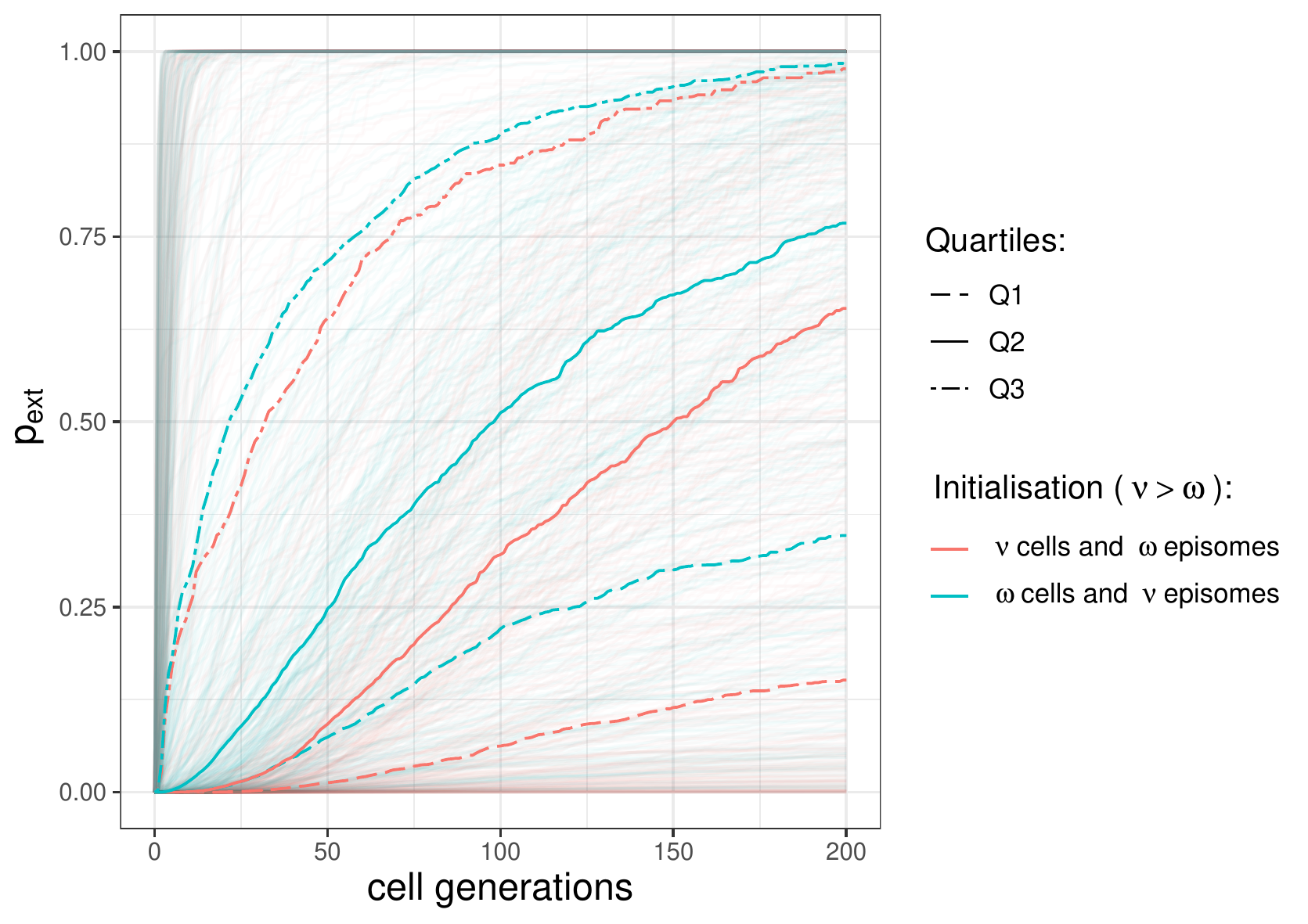}
    \caption{\textbf{Estimation of the cumulative probabilities of extinction and their quartiles for the two extreme scenarios.} In red are displayed the estimations when the $N_0\times\Gamma_0$ initial viral copies are allocated in $\nu$ cells, each containing $\omega$ episomes whereas in blue, the episomes are distributed in $\omega$ cells, each containing $\nu$ copies ($\nu>\omega$). The three quartiles are lower for the first scenario, thus indicating a lower risk of extinction when the copies are split in the maximum number of cells possible.}
    \label{fig:comp_pext_nu}
\end{figure}

To determine which scenario favours virus persistence we compare the cumulative probabilities of extinction for each of pairs of parameters that only differ by the initial arrangement of the viral copies. Figure \ref{fig:comp_pext_nu} shows the quartiles for the two extreme scenarios. This indicates a benefit for the virus to spread its copies to the maximum number of stem cells at the start of the infection, the bigger the difference between $\nu$ and $\omega$ the more significant this benefit is. 

\FloatBarrier
\section{Discussion}

As immune mechanisms failed to answer completely the problem of spontaneous clearance in natural infections, \cite{Ryser2015} has suggested random events in the type of cell divisions could induce extinction and thus complement explanation based on immunity. Following this approach, we refine the model initially exposed by \cite{Ryser2015} and emphasize the risk of stochastic extinction. We hypothesized that an HPV infection could face extinction due to the sole effects of random processes occurring both at the intra- and inter-cellular level. Using a stochastic model comprised of two nested branching processes and numerical simulations, we find that random allocation of episomes during cell divisions drastically limit the persistence in the early phase of the epidemic. Later, if episomes persisted in some infected cells, randomness in the type of cell divisions can also lead to extinction, even though depending on the local context, symmetric divisions can also favour the spread of the virus in the tissue (when $s>r$). The importance of each aspect varies over throughout the infection, and more generally as HPV viral copies increase in number, the role of stochasticity in the explanation of extinction fades. Our results show the first stages of the infection matter a lot for the persistence of the virus, which is particularly important for viruses like HPV that have a  low number of copies initially \cite{Doorbar2012}.

Another outcome of our model is that many infections by HPV might clear naturally in a matter of weeks/months before even being noticed by the host immune system since episomes remain in small number in the basal layer. Adding to the pioneering work of \cite{Ryser2015} to highlight  the role of stochastic processes in the clearance of HPV infections, we show that adding intracellular stochasticity greatly impacts extinction probability, especially in the early stages of the infection. Combining the two processes enables us to reach susbtantial clearance rates  without involving immune response.

These results could explain why the antibody production against the virus is low after natural infections \cite{Stanley2008, Doorbar2012}, as the infection might just fade randomly in within the first divisions of the initially infected cells. HPVs might have evolve to limit pressures exerted by randomness during the first cell divisions by amplifying the viral copies in the early phase of the epidemic, notably with the expression of E1 protein \cite{Egawa2012}, until it reaches maximum host cell capacity. The same report showed that the E1 protein was decisive in the establishment of a persistent infection but, once the number of episomes per cell reaches host capacity, E1 protein was dispensable for the maintenance of the infection. Our model corroborates these observations: we show that the early phase of the infection can be detrimental to the infection, E1 protein could favour the maintenance of the infection either by acting on the probability to be distributed to the daughter stem cell ($p$ parameter) or promote episomes amplification ($\lambda$ parameter). Upon reaching host cell capacity, even if the intracellular regime is critical, it is very unlikely for the virus to go extinct, hence E1 protein is dispensable during maintenance phase of the infection \cite{Egawa2012}.

Globally our results suggest that remaining with low copy number is a way to evade immunity \cite{Doorbar2013}, but such a strategy is risky for the virus is under strong stochastic forces that plummet the persistence of the infection. Viruses committed to such strategy might alleviate these pressures exerted by randomness using proteins in the very early phase of the infection, like the E1 protein for HPV. Yet, such proteins might not be the most adequate targets for antiviral drugs as papers showed that the role of these elements are crucial in the very early steps of the infection but lose importance once host capacity is reached \cite{Egawa2012}. We might not be able to detect HPV sufficiently early to inhibit E1 protein to cure infections.

\cite{Ryser2015} assume a balance between the two types of symmetric divisions, thus restraining the inter-cellular dynamic to the critical regime. We allow the possibility for the tissue to deviate from that assumptions.
Notably, we allow the probability of symmetric divisions in two stem cells to exceed the probability of symmetric divisions in two differentiated cells as keratinocytes can alternate between two modes of proliferation depending on local context. If local confluence is removed by scratch injury, cells switch to the expanding mode, characterized by an excess of cycling cells production, until confluence is reached again, after which they switch back to a balanced mode where similar proportions of proliferating and differentiating cells are produced, thus generating population asymmetry maintaining epidermal homeostasis \cite{Roshan2016}. The two modes of proliferation could differ by the rates of cell divisions or the probability of division in two stem cells (parameter $s$), or both.
Our model predicts that HPVs greatly benefit from the expanding mode, which echoes previous empirical work on the link between HPV persistence and wound healing responses \cite{Ledwaba2004, Doorbar2006,Fuchs2008, Schiller2010}. 

To further calibrate the models and delineate the true role of stochasticity, a comparison with longitudinal follow-ups of women recently infected by any type of HPV or presenting chronic latent infections would be of primer interest. Ideal time windows between two measurements should not exceed 1 month. Besides, we should perhaps reconsider how we label infections that cleared in less than 1 month. Many follow-ups would treat them as "carriage", i.e. HPV from the partner and not a "true" infection. Perhaps there are more non-persistent infections than we thought? Overall to decipher this issue, we need accurate infection duration distributions over short time scales, ideally with virus loads measures.

Comparing longitudinal follow-ups data and numerical simulations would require clarifying the estimation of the time between two cell divisions. We rely on old measures \cite{Averette1970} to set the upper bound of the number of cell generations to compute in our three years simulations. This work estimates the rate of divisions around [0.03, 0.07] day\up{-1}. These results are coherent with more recent work on the kinetics of epidermal maintenance stating an overall division rate of $1.1$ per week \cite{Klein2007}. To fit his model to follow-ups data, \cite{Ryser2015} estimated the rate of division using proxies (height of the stratified squamous epithelium expressed in number of cells, renewal time of the tissue, fraction of proliferative cells in the basal layer) and find a much higher division rate [0.29,1] day\up{-1}. Updates on the estimation of this variable would be of primary interest to compare theoretical and experimental approaches.

\section{Acknowledgements}

The authors acknowledge the IRD itrop HPC (South Green Platform) at IRD Montpellier for providing HPC resources that have contributed to the research results reported within this paper. TB is funded by a doctoral fellowship from the Ligue Contre le Cancer. SA acknowledges the support from European Research Council (ERC) under the European Union’s Horizon 2020 research and innovation program (EVOLPROOF, grant agreement No 648963). Further support was provided by the CNRS, the IRD and the University of Montpellier.

\newpage


\newpage

\appendix

	\renewcommand{\thetable}{S\arabic{table}}
	\renewcommand{\thefigure}{S\arabic{figure}}
	\renewcommand{\thesection}{S\arabic{section}}
	\renewcommand{\theequation}{S\arabic{equation}}
	
	\setcounter{figure}{0}
	\setcounter{table}{0}
	\setcounter{section}{0}
	\setcounter{equation}{0}

\section{Details on intracellular branching processes}
\label{appendix:intra_process}

We distinguish two cases, constant or Poisson amplification, for the two processes are slightly different. We remind that $X_n$ is a random variable following the total number of episomes in all infected stem cells. Conditioned on the absence of symmetric cell divisions, we model the variation of $X_n$ over time as a Bienaymé-Galton-Watson (BGW) branching process defined by the following recurrence relation, for all $n\in\mathbb{N}$

\begin{align}
    X_{n+1} = \sum\limits_{i=1}^{X_n} \xi_i 
\end{align}

where $\xi_i=\xi$ are independent and identically distributed random variables modeling the offspring number of a single individual within a generation. 

\paragraph{Dirac episome multiplication}

In this case, each episomes distributed to the daughter stem cell gives exactly $\lambda$ new episomes then dies, so the random variable $\xi$ can be written as

\begin{align}
    \xi = \left\{ \begin{array}{cc}
         0 & \text{with probability $1-p$} \\
         \lambda &  \text{with probability $p$}
    \end{array}
    \right.
\end{align}
Hence we deduce the associated probability generating function (PGF) $g(z), z\in[0,1]$:

\begin{align}
    g(z) =  \sum \limits_{k=0}^{\infty} \mathbb{P}[\xi = k]\,z^k=  1-p + p \, z^\lambda
\end{align}

Following classical results on BGW branching processes \cite{Athreya1972},  $\mathbb{P}[X_\infty = 0|X_0 = 1] = p_{\text{ext}}$ is the smallest fixed point of the PGF on the interval $z\in[0,1]$: 
$$p_{\text{ext}} = \inf \{z \in[0,1], g(z) = z\}$$
One can show that if $m = g'(1) = \lambda\,p\leq 1$, the average number of viral copies produced by one episome between each iteration,  then $p_{\text{ext}}=1$. If $m>1$ one must solve the following polynomial equation to find $p_{\text{ext}}$

\begin{align}
    1-p -z + p z^\lambda = 0
\end{align}

\begin{figure}[ht]
    \centering
    \includegraphics{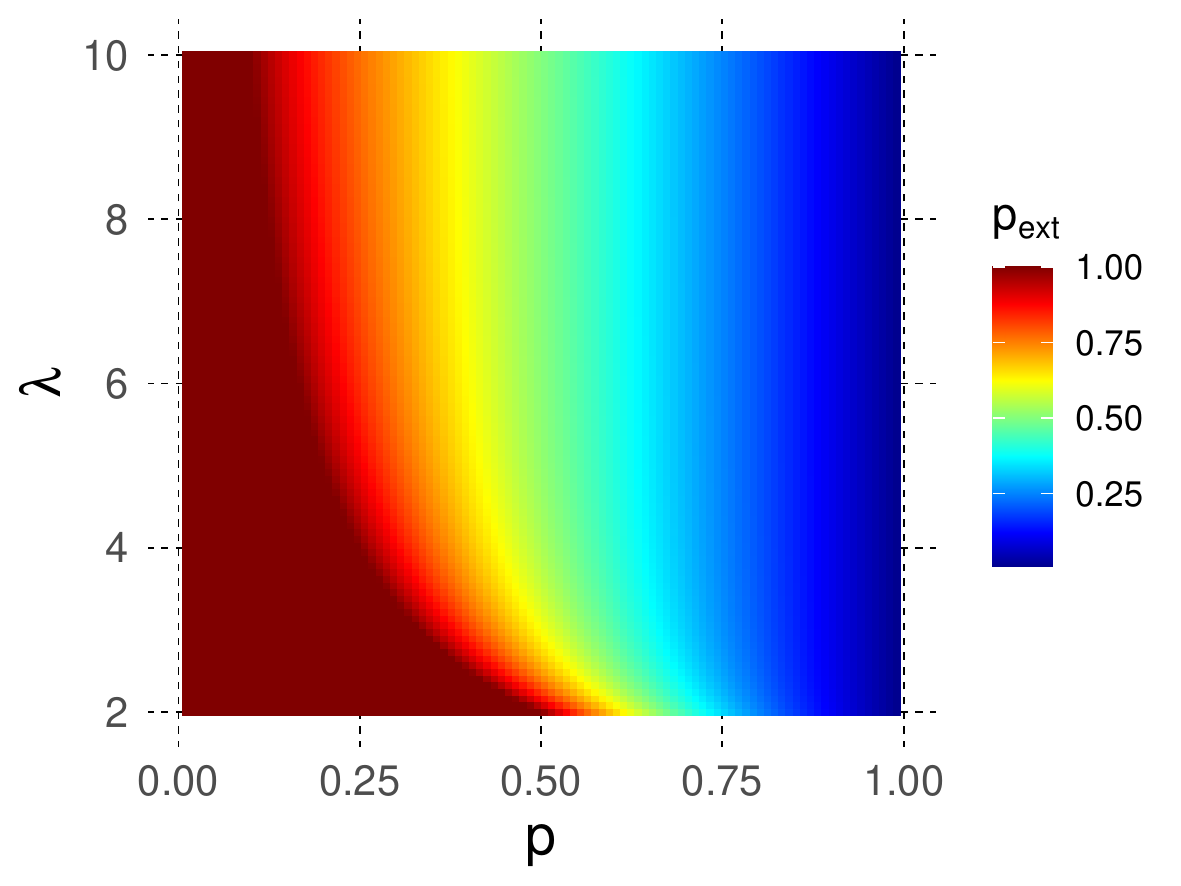}
    \caption{\textbf{Effect of episome replication ($\lambda$) and asymmetry in episome partitioning ($p$) on the cumulative probability of extinction ($p_{\text{ext}}$).} We here assume a Dirac scenario. $p_{\text{ext}}$ decreases with both $\lambda$ and $p$, and reaches a threshold $p_{\text{ext}}=1-p$ for a given $p$ when $\lambda$ increases. This is consistent with the fact that if $\lambda$ is sufficiently high, the main source of extinction is the first cell division of a stem cell containing 1 episome.}
    \label{fig:pext_cst_post}
\end{figure}

For $\lambda\in\{2,3\}$, such solutions are trivial but for higher value of $\lambda$, the solution becomes intractable. The figure \ref{fig:pext_cst_post} shows the variation of extinction probability with the parameters $\lambda$ and $p$.

\FloatBarrier

\paragraph{Poisson episome multiplication}

We now assume that upon division, each episome dies and gives $\rho$ descendants, where $\rho$ is a random variable following a Poisson distribution, $Poisson(\lambda)$. Hence the distribution of $\xi$ follows:

\begin{align}
    \mathbb{P}[\xi = k] = \mathbb{1}_{\{k=0\}} (1-p) + p \,\mathcal{P}^\lambda(k)
\end{align}

where $\mathcal{P}^\lambda(k)$ is the probability mass function of the Poisson distribution $Poisson(\lambda)$. The PGF associated with such process is 

\begin{align}
    g(z) = 1-p+p\,e^{\lambda(z-1)}
\end{align}

Similarly to the previous case, if $m=\lambda\,p \leq1$, the probability of extinction is equal to 1, else this probability is lower than 1. Let $W$ be the Lambert function defined as the inverse function of $z\mapsto ze^z$. If $m>1$ we have

\begin{align}
    p_{\text{ext}} = \frac{-W(-\lambda\,p\,e^{-\lambda\,p}) + \lambda\,(1-p)}{\lambda}
\end{align}

The sensitivity of $p_{\text{ext}}$ with the parameters $\lambda$ and $p$ are displayed in figure \ref{fig:pext_poiss_post}. Between the two cases, there are little differences, except for low value of $\lambda$ where $\mathcal{P}^\lambda(0)$ is non-negligible, hence favoring extinction compared to the Dirac case. 



\begin{figure}[ht]
    \centering
    \includegraphics[scale = 0.75]{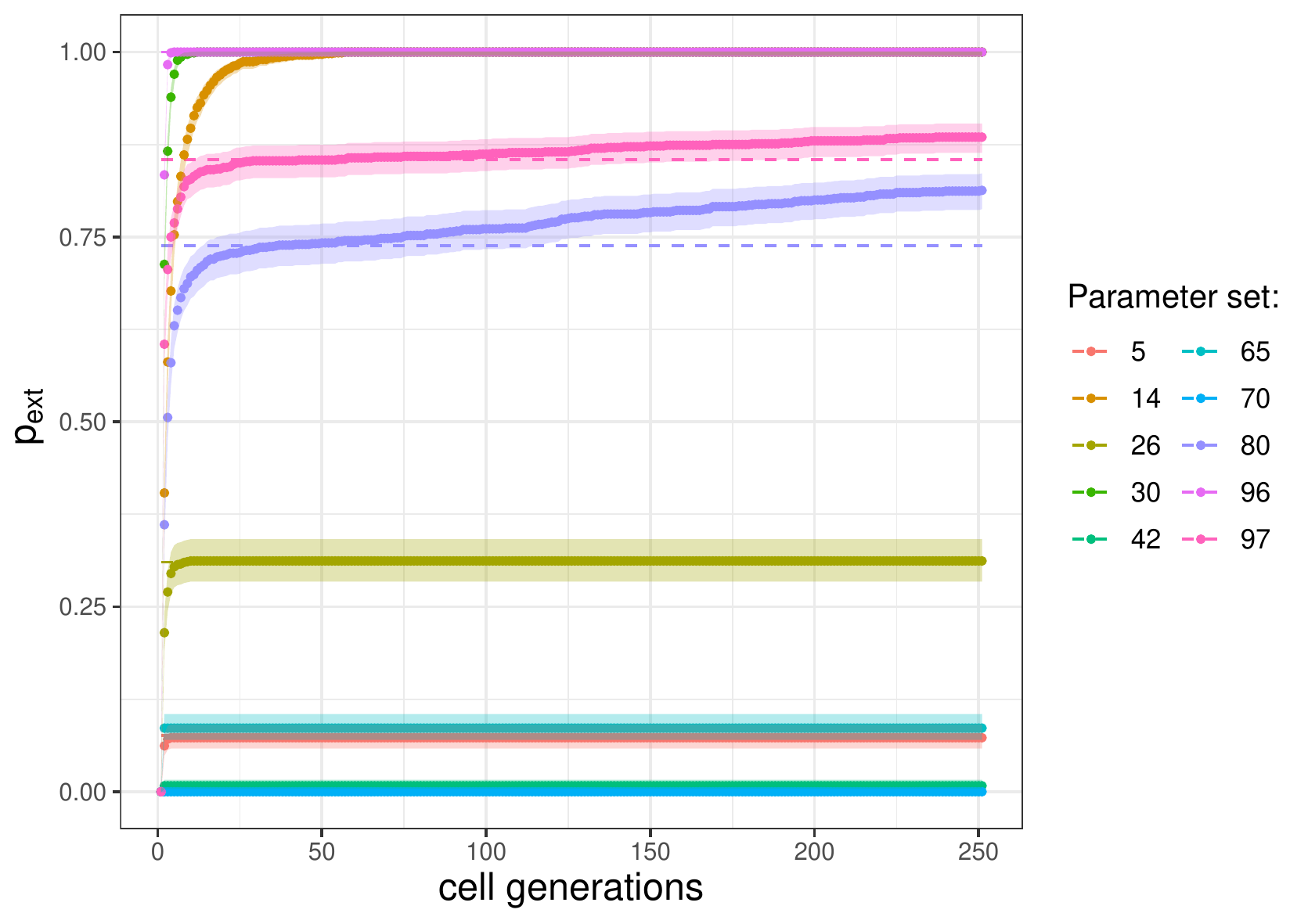}
    \caption{\textbf{Estimated cumulative probability of extinction ($p_\text{ext}$ with finite intra-host capacity (plain dots, ribbons correspond to $95\%$ confidence intervals) and according to theory (dashed lines).} In most cases, the estimations of $p_\text{ext}$ under constrained host capacity do not diverge from theoretical predictions on the time window considered. Deviations emerge when the intra-host regime is slightly supercritical and/or the intra-host capacity limited (<50). The colors indicate the parameters used for each simulation: the number indicates the row of the parameter matrix \textsf{LHS\_intra} (see supplementary materials).}
    \label{fig:fit_intra}
\end{figure}

\FloatBarrier
\section{Intracellular dynamics: amplification step prior to distribution step}

\label{appendix:amplification_before_partition}

Assume we reverse the two steps of the intracellular described in \ref{section:intra_cell_dyn}. Upon division, the episomes are first amplified before being distributed in the two daughter cells. If we put the time origin just after the first amplification phase, we notice we are back in the scenario where the distribution of the viral copies occurs before the amplification phase. Therefore we deduce that a process starting with amplification phase is equivalent to the same process starting with distribution step and extra initial viral copies. From classical results on BGW branching process, we have the following results, for all $\nu\in\N^*$:

\begin{align}
    \mathbb{P}[X_\infty = 0|X_0 = \nu] = \left(\mathbb{P}[X_\infty = 0|X_0 = 1]\right)^\nu
\end{align}

Let $p_{\text{ext}}^D$ be the probability of extinction when the intracellular process starts with the distribution step and $p_{\text{ext}}^A$ the probability of extinction when it begins with the amplification phase. For all $\lambda\in\mathbb{N}$ we hence have:

\begin{align}
    \left(p_{\text{ext}}^D\right)^\lambda = p_{\text{ext}}^A
\end{align}

\section{Detail on the probability of extinction in the symmetric division case.}

\label{appendix:inter_process}

We remind the inter-cellular process is modeled using the following discrete BGW branching process. Let $\Gamma_n$ be the random variable describing the total number of infected stem cell at time $n\in\N$. We note by $\widetilde{\Gamma}_n$ the random variable describing the dynamic of infected stem cells when the number of episomes in each stem cell is equal to $\infty$ (i.e. when omitting intracellular stochasticity). This variable is defined by the following recurrence relation:

\begin{align}
    \widetilde{\Gamma}_{n+1} = \sum\limits_{i=1}^{\widetilde{\Gamma}_n} \phi_i
\end{align}

where $\phi_i=\phi$ are independent and identically distributed random variables modeling the offspring number of a single infected cell within a generation. From equation \ref{eq:branching_general} we can characterised $\phi$ as follow

\begin{align}
    \phi = \left\{ \begin{array}{cl}
        2 &  \text{with probability $s$}\\
        0 &  \text{with probability $r$}\\
        1 &  \text{with probability $1-r-s$}
    \end{array}
    \right.
\end{align}

For all $z\in[0, 1]$, the PGF $h(z)$ associated to this reproduction law is defined as:

\begin{align}
    h: z \to r + (1-r-s)\,z + s\,z^2
\end{align}

We deduced that the average number of stem cell produced by each dividing stem cell is equal  to $m=1-r+s$. The probability of extinction $p_{\text{ext}}$ follows

\begin{align}
    p_{\text{ext}} = \inf \{ z \in[0, 1], h(z) = z\}
\end{align}

In this case the results is straightforward and $p_{\text{ext}}$ is the smaller roots of the polynomial equation 

\begin{align}
     r -(r+s)\,z + s\,z^2 = 0, \,z\in[0,1]
\end{align}

Hence we deduced $p_{\text{ext}} = \min \{ \frac{r}{s}, 1\}$

\begin{figure}[ht!]
    \centering
    \includegraphics[scale = 0.75]{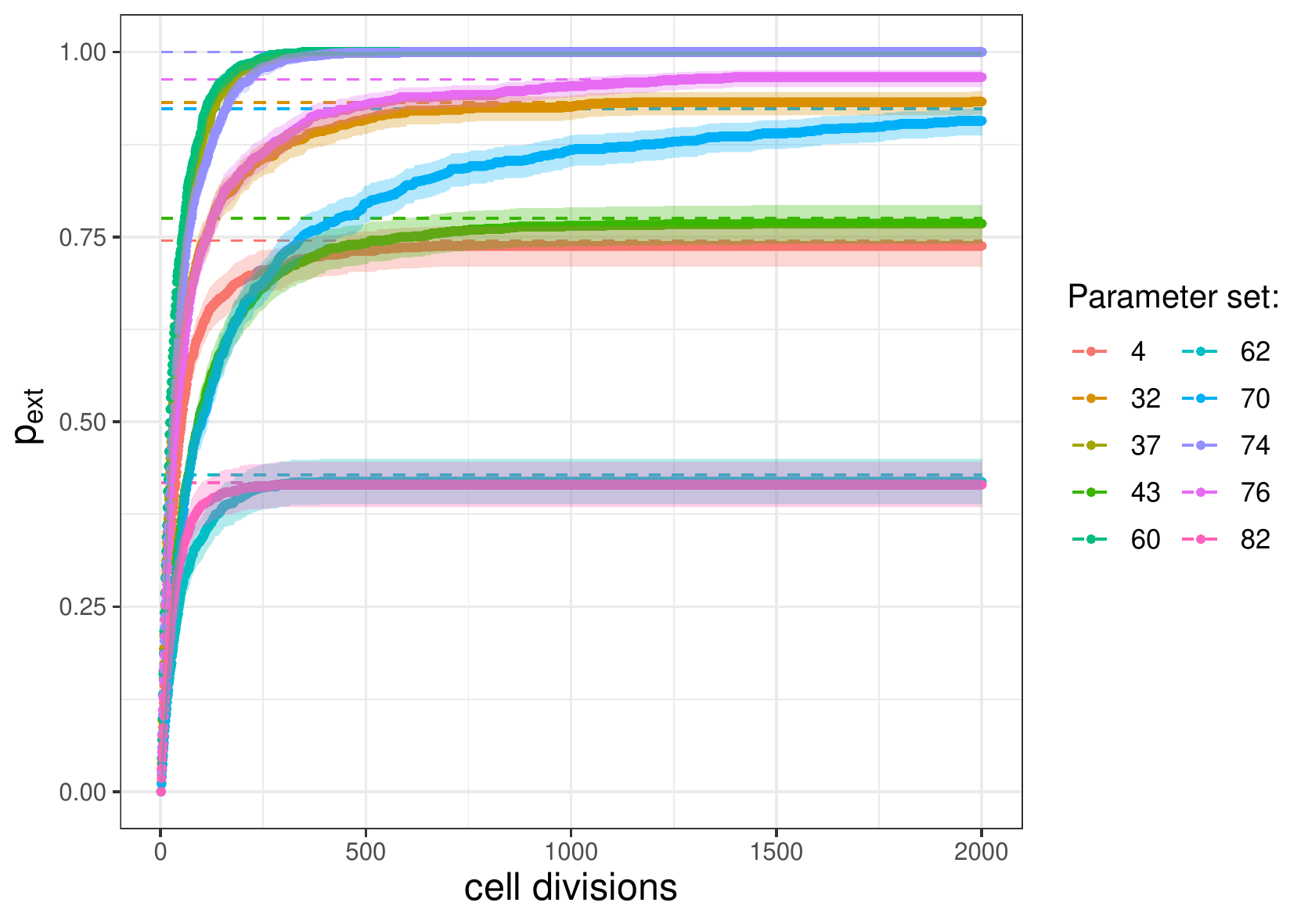}
    \caption{\textbf{Estimated cumulative probability of extinction ($p_\text{ext}$ with finite intra-host capacity (plain dots, ribbons correspond to $95\%$ confidence intervals) and according to theory (dashed lines).} We observe no significant differences between estimations and theoretical predictions. The colors indicate the parameters used for each simulation: the number indicates the row of the parameter matrix \textsf{LHS\_inter} (see supplementary materials).}
    \label{fig:fit_inter}
\end{figure}
\FloatBarrier
\FloatBarrier
\subsection*{Lower bound of the probability of extinction}

We remind that $\Gamma_n$ (resp. $\widetilde{\Gamma}_n$) is the random variable describing the dynamic of infected stem cells over time in the general framework (resp. in the absence of intracellular stochasticity) at time $n$ and $X_n^{(k)}$ the random variable characterizing the number of viral copies in lineage $k$ at time $n$. We can write that:

\begin{align*}
   \mathbb{P}[\Gamma_{n+1}=0|\Gamma_n = \alpha] &= \mathbb{P}[\widetilde{\Gamma}_{n+1}=0|\widetilde{\Gamma}_n = \alpha] + \sum\limits_{i=1}^\alpha \mathbb{P}[\widetilde{\Gamma}_{n+1}=i|\widetilde{\Gamma}_{n} = \alpha] \times \sum\limits_{k=1}^i \mathbb{P}[X_{n+1}^{(k)} = 0|X_{n}^{(k)}>0]\\
    \mathbb{P}[\Gamma_{n+1}=0|\Gamma_n = \alpha] & \geq \mathbb{P}[\widetilde{\Gamma}_{n+1}=0|\widetilde{\Gamma}_n = \alpha]
\end{align*}

This relation holds for all $n\in\N$, thus we deduce that
\begin{align}
    \mathbb{P}[X_\infty=0|\Gamma_0=1] \geq \min\left(\frac{r}{s}, 1\right) 
\end{align}.

\section{Parameters generations using LHS}
\label{appendix:LHS}
We generate two matrixes of parameters using R package \og lhs\fg \cite{lhsR}. This package generates matrixes $M = (m_{i,j})_{(1 \leq i \leq n, 1\leq j \leq k)}$ where
\begin{align*}
    0\leq m_{i,j} \leq 1 \quad ; \quad 1 \leq i \leq n,  1\leq j \leq k
\end{align*}

As most of our parameters are defined outside of $I=[0,1]$, we modify $(m_{i,j]})$ so that we get meaningful results regarding the parameter support. Let $[a_j, b_j]$ be the definition domain of variable $j$, $a_j\leq b_j$, then the value of parameter $j$ for the $i$-th set, $S_{i,j}$ is  

\begin{align}
    S_{i,j} = a_j + m_{i,j} (b_j-a_j)
\end{align}

In the case where variable $j$ is an integer we rounded up $S_{i,j}$. Concerning the choice of the intracellular multiplication regime (Dirac or Poisson distributed), we assign the Fixed multiplication regime when $(m_{i,j}) <0.5$ and the Poisson regime otherwise. See Table \ref{tab:Notation} for more details on the domain of definition of numerical variables. Additionally we set the upper bound for $\lambda$ and $N_0$ to 10.

\section{Annotation of infected cell lineages}

\label{appendix:annotation}

Similarly to phylogeny, our process viewed from the infected stem cell perspective, can be visualised as a tree whose origin is the first infected cell. As long as the cells descending from that first infected cell divide asymmetrically, we remain on the same branch: we group such cells in a cell lineage. The moment, a stem cell divides symmetrically in two stem cells, the tree is split into two new branches that follow the same process independently. When a symmetric divisions into two differentiated cells occurs, we stop the branch. To track each branch we use the following notations.

\begin{enumerate}[label=\roman*)]
    \item The first cell lineage is labelled \textbf{O}. 
    
    \item The two daughter stem cells are labelled \textbf{L} and \textbf{R} from its parent cell point of view. To distinguish them from other stem cells in the tree, they are named as follow:
    \begin{itemize}
        \item Take the parent's name (usually the letter \textbf{O} followed by a succession of \textbf{L} or \textbf{R}).
        \item Add the letter \textbf{L} or \textbf{R} to the parent's name.
    \end{itemize}
\end{enumerate}

For instance the two new stem cells lineage descending from the initial infected stem cell lineage are labelled \textbf{OL} and \textbf{OR}. Or, upon division, the stem cell lineage \textbf{OLLR} gives two new stem cell lineages \textbf{OLLRL} and \textbf{OLLRR}. The choice of $\textbf{L}$ or $\textbf{R}$ is random and have no influence on the rest of the process.



\section{Simulation of our process}
\label{appendix:simu}

As the intracellular branching process is nested inside the inter-cellular branching process, we first need to compute the underlying tree of infected cells, then simulate the trajectories of the viral copies for each cell lineage (i.e. each branch) starting from the origin. We simulated the tree and the different trajectories using Python v3.9.1+. 

\subsubsection*{Determination of the inter-cellular tree}

The branching process for the inter-cellular dynamic can be viewed as a discrete birth-death process. Each individual, upon division, randomly takes one of the following three paths: i) divide symmetrically in two differentiated cells (death) with probability $r$ ii) divide symmetrically in two stem cells (birth) with probability $s$ iii) divide asymmetrically (pursue its life) with probability $1-r-s$. Thus for each individual we can compute the time of symmetric divisions as it follows geometric distribution $Geometric(r+s)$. Once the time of symmetric division is determined we randomly assign of the two outcomes by drawing a random number from uniform distribution $Uniform([0,1])$: if this number is lower than $r/(r+s)$ then the division gives two differentiated cells, otherwise the cell divides into two stem cells. We repeat this method until all cells have divided into two differentiated cells or when the remaining cell divisions time occur after a given threshold.

\subsubsection*{Intracellular dynamic for each branch}
\label{appendix:stochastic_matrix}

The intracellular branching process can also be consider as a Markov-chain. To simulate the dynamic of a Markov-chain we rely on its associated stochastic Matrix $P=(P_{i,j})_{(i,j)\in\N^2}$, where $P_{i,j} = \mathbb{P}[X_1=j|X_0=i]$. Depending on the episomes amplification regime (Fixed of Poisson - amplification or allocation phase first), the matrix varies. We detail the matrix $P$ for the Fixed and Poisson cases when allocation occurs before amplification. We remind that when amplification occurs before allocation, the results on the probabilities of extinction are similar to the result obtained when the two phases are reversed raised to the power $\lambda$ (see Appendix \ref{appendix:amplification_before_partition}). First we explicit the stochastic matrix in the Fixed case:

\begin{align}
   P_{i,j} = \left \{
   \begin{array}{cl}
        \mathcal{B}^i_p(j/\lambda) & \text{if } j\in \lambda\mathbb{N}\; ;\; j\leq\lambda\,i < C \\
       1-\sum\limits_{k=0}^{C-1} P_{i,k} & \text{if $j=C$}\\
       0 & \text{otherwise}
   \end{array}
   \right .
\end{align}

We note by $\mathcal{B}^n_p(x)$ the probability of drawing $x$ individuals from a binomial distribution $Binomial(n, p)$. If we now assume a Poissonian amplification regime with distribution $Poisson(\lambda)$, then for all $(i,j)\in\N^2$:

\begin{align}
        P_{i,j} = \left \{ \begin{array}{cll}
             \sum \limits_{k=0}^{i} \mathcal{B}^i_p(k)\,\mathcal{P}^{\lambda k}(j) & \text{if} & j < C \\
              1-\sum\limits_{k=0}^{C-1} P_{i,k} & \text{if} & j = C\\
              0 & \text{otherwise}
    \end{array}  \right.
    \end{align}
We note by $\mathcal{P}^{n}(k)$ the probability of drawing $k$ individuals from a Poisson distribution $Poisson(n)$. As the state $0$ is included, $P$ is squared matrix of size $C+1\times C+1$. Given the stochastic matrix $P$ for all iteration $n\in\N$ and $i\in\llbracket0, C\rrbracket$, we iteratively determine the trajectory between two events of symmetric divisions by randomly choosing the next value for iteration $n+1$

\begin{figure}[ht]
    \centering
    \includegraphics[height = 7cm, width = 12cm]{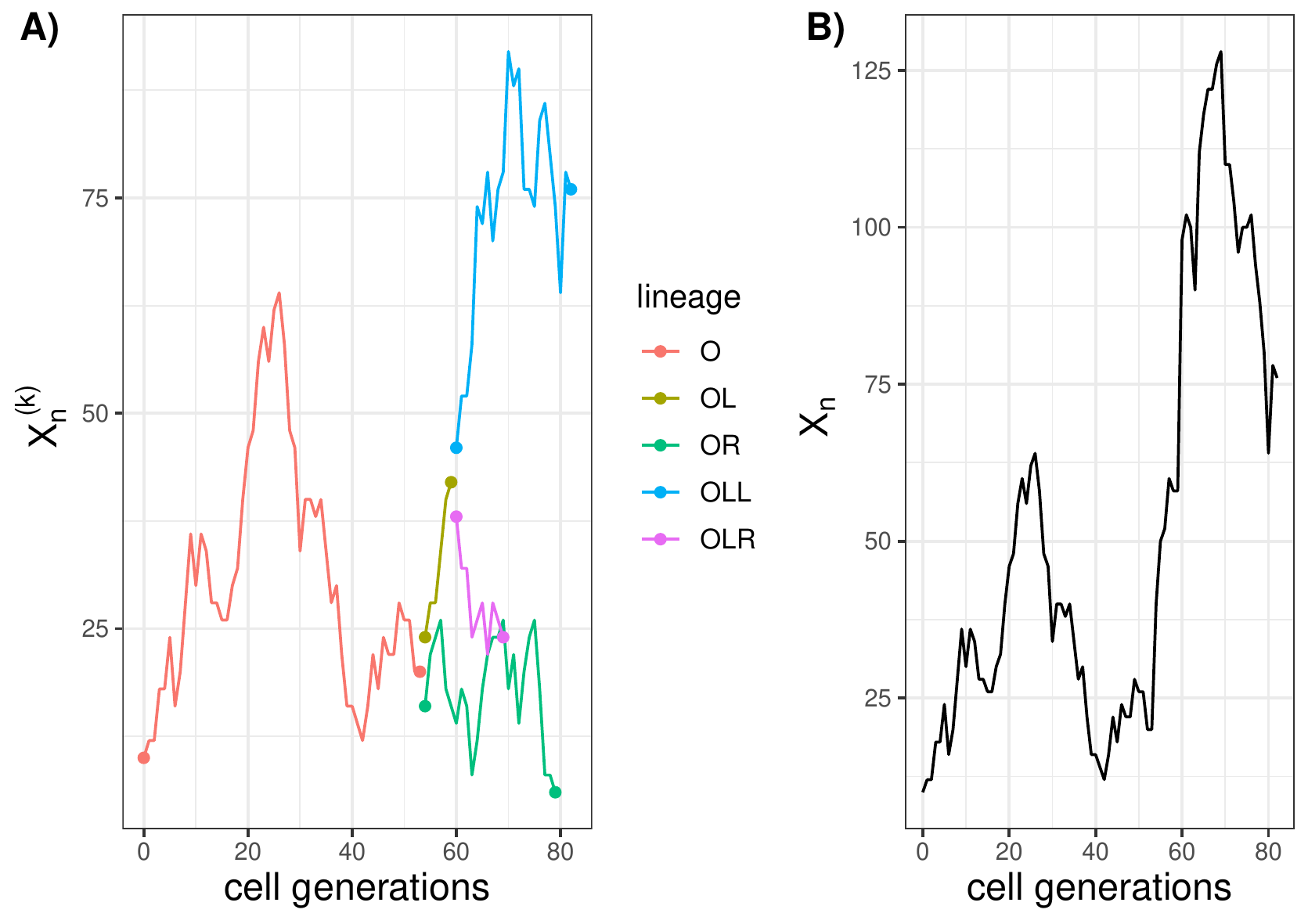}
    \caption{\textbf{Stochastic trajectory of the dynamic of the number of episomes over time} On the left panel (A) we displayed the dynamics inside each cell lineage over time. On the right panel (B) we plot the variation in the total number of episomes.
    We fixed the value of the parameters as follows: $p=0.5$, $\lambda = 2$, $s = r = 0.02$, $N_0 = 10$ and $C=200$. The episomes amplification is fixed and happens after cell divisions. On panel (A), each color represents a cell lineage (see appendix \ref{appendix:annotation} for more information on cell lineage labelling).}
    \label{fig:simu1}
\end{figure}
\FloatBarrier

\FloatBarrier

\section{Effects of the initial conditions}
\begin{figure}[ht]
    \centering
    \includegraphics[scale = 0.8]{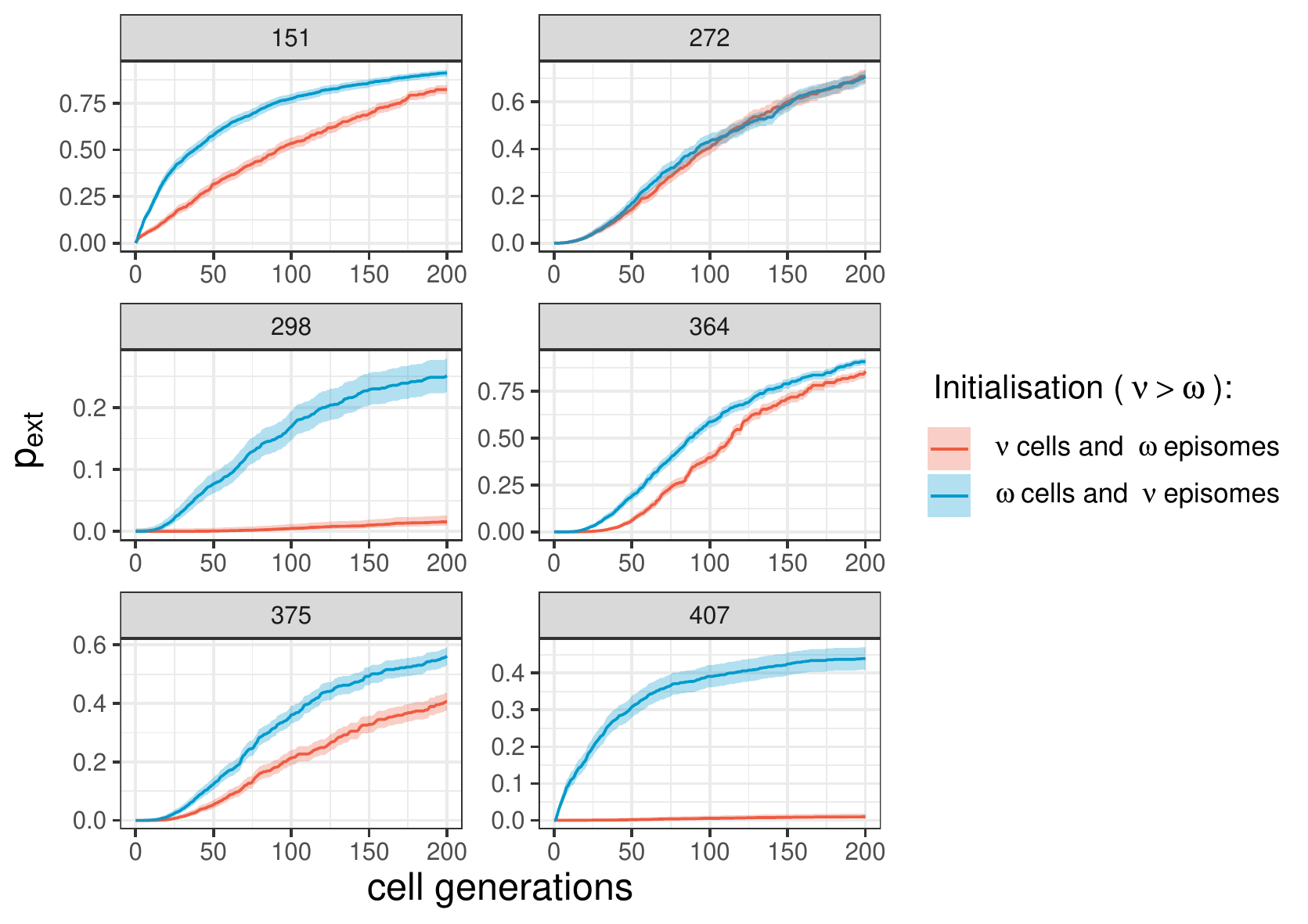}
    \caption[list=off]{\textbf{Estimations of the cumulative probability of extinction ($p_\text{ext}$) for two sets of parameters that only differ by the allocation of the viral copies at the start of infection.} We display in red the estimations of $p_\text{ext}$ in the scenario where the initial few viral copies are spread in more cells and in blue the scenario where more viral copies are spread in less cells. The solid lines indicates the estimations of $p_\text{ext}$ and the ribbon the 95\% confidence intervals. The cumulative probability of extinction is generally lower in the first scenario compared to the latter. Thus it is more beneficial to spread its copies in the maximum number of cells at the the start of the infection. On each facet, the title indicates the row of the parameter matrix \textsf{LHS\_1} (see supplmentary materials).}
    \label{fig:pair_nu}
\end{figure}
\end{document}